\documentclass[reprint,amsmath,aps,prl,twocolumn]{revtex4-1}
\usepackage[utf8]{inputenc}
\usepackage[T1]{fontenc}
\usepackage{graphicx}
\usepackage{dcolumn}
\usepackage{amsmath}
\usepackage{amssymb}
\usepackage{array}
\usepackage{multirow}
\usepackage{calc}
\usepackage{color}
\usepackage{graphicx}
\usepackage{float}
\usepackage{bm}
\usepackage{hyperref}

\newcommand{\icm}{\ensuremath{~\textrm{cm}^{-1}} }

\begin{document}

\title{Symmetry of the Excitations in the Hidden Order State of URu$_2$Si$_2$}
\author{J. Buhot}
\affiliation{Laboratoire Mat\'eriaux et Ph\'enom\`enes Quantiques, UMR 7162 CNRS, Universit\'e Paris Diderot, B$\hat{a}$t. Condorcet 75205 Paris Cedex 13, France}
\author{M.-A. M\'easson}
\email{marie-aude.measson[at]univ-paris-diderot.fr}
\affiliation{Laboratoire Mat\'eriaux et Ph\'enom\`enes Quantiques, UMR 7162 CNRS, Universit\'e Paris Diderot, B$\hat{a}$t. Condorcet 75205 Paris Cedex 13, France}
\author{Y. Gallais}
\affiliation{Laboratoire Mat\'eriaux et Ph\'enom\`enes Quantiques, UMR 7162 CNRS, Universit\'e Paris Diderot, B$\hat{a}$t. Condorcet 75205 Paris Cedex 13, France}
\author{M. Cazayous}
\affiliation{Laboratoire Mat\'eriaux et Ph\'enom\`enes Quantiques, UMR 7162 CNRS, Universit\'e Paris Diderot, B$\hat{a}$t. Condorcet 75205 Paris Cedex 13, France}
\author{G. Lapertot}
\author{D. Aoki}
\affiliation{Univ. Grenoble Alpes, INAC-SPSMS, F-38000 Grenoble, France}
\affiliation{CEA, INAC-SPSMS, F-38000 Grenoble, France}
\author{A. Sacuto}
\affiliation{Laboratoire Mat\'eriaux et Ph\'enom\`enes Quantiques, UMR 7162 CNRS, Universit\'e Paris Diderot, B$\hat{a}$t. Condorcet 75205 Paris Cedex 13, France}

\begin{abstract}

We have performed polarized electronic Raman scattering on URu$_2$Si$_2$ single crystals at low temperature down to 8~K in the hidden order state and under magnetic field up to 10~T.
The hidden order state is characterized by a sharp excitation at 1.7~meV and a gap in the electronic continuum below 6.8~meV. Both Raman signatures are of pure A$_{2g}$ symmetry. By comparing the behavior of the Raman sharp excitation and the neutron resonance at \textbf{Q$_0$}=(0,0,1), we provide new evidence, constrained by selection rules of the two probes, that the hidden order state breaks the translational symmetry along the c axis such that $\Gamma$ and $Z$ points fold on top of each other. The observation of these distinct Raman features with a peculiar A$_{2g}$ symmetry as a signature of the hidden order phase places strong constraints on current theories of the hidden order in URu$_2$Si$_2$.

\end{abstract}

\date{\today}

\maketitle

For almost three decades \cite{palstra_superconducting_1985}, the identity of the ordered phase found in URu$_2$Si$_2$ at temperature below T$_0$=17.5~K has eluded researchers \cite{mydosh_colloquium:_2011, mydosh_hidden_2014} despite intensive experimental and theoretical investigations. The second order transition to this so-called hidden order (HO) appears clearly in the thermodynamic and transport quantities \cite{palstra_superconducting_1985, schlabitz_superconductivity_1986, behnia_thermal_2005, schoenes_hall-effect_1987, de_visser_thermal_1986}. This unique electronic state is not a simple long-range magnetic (dipolar) order since the associated tiny magnetic moment measured in the HO state cannot account for the large entropy release during the transition \cite{broholm_magnetic_1991}. Nevertheless, the HO changes to a simple antiferromagnetic with a simple tetragonal structure under a small applied pressure of 0.5~GPa \cite{amitsuka_pressuretemperature_2007, butch_antiferromagnetic_2010, hassinger_temperature-pressure_2008}.
Many interesting theories have been proposed to explain the nature of HO, among which multipolar orders from quadrupolar to dotriacontapolar \cite{kusunose_hidden_2011-1, ressouche_hidden_2012, haule_arrested_2009, ikeda_emergent_2012, rau_hidden_2012}, local currents \cite{chandra_hidden_2002, fujimoto_spin_2011}, unconventional density wave \cite{riseborough_phase_2012, das_imprints_2014} modulated spin liquid \cite{pepin_modulated_2011, thomas_three-dimensional_2013}, dynamical symmetry breaking \cite{elgazzar_hidden_2009} and hastatic order \cite{chandra_hastatic_2013} for the most recent ones. Yet a complete understanding of the nature of the hidden order has not been reached.
\par
A wide variety of experimental studies have succeeded in revealing salient features of the HO state. Inelastic neutron measurements \cite{broholm_magnetic_1991, bourdarot_precise_2010, wiebe_gapped_2007} observe two magnetic excitations with a commensurate wave vector Q$_0$=(1,0,0)$\equiv$(0,0,1) and an incommensurate wave vector Q$_1$=(1.4,0,0) at 1.7~meV and 4.8~meV, respectively. The first one has been demonstrated to be a major signature of the HO state \cite{villaume_signature_2008}. It is well accepted \cite{bonn_far-infrared_1988, schoenes_hall-effect_1987} that a partial Fermi-surface gapping with a strong reduction of the carriers number occurs at T$_0$ and accordingly, the electronic structure abruptly reconstructs at T$_0$ \cite{santander-syro_fermi-surface_2009,yoshida_signature_2010,boariu_momentum-resolved_2013,meng_imaging_2013}. It persists in the antiferromagnetic state under pressure \cite{hassinger_similarity_2010} suggesting similar Brillouin zone folding in both states. Besides, a recent set of experiments have identified a four-fold symmetry breaking upon entering the HO state  \cite{okazaki_rotational_2011, tonegawa_cyclotron_2012, tonegawa_direct_2014}, steering some controversy \cite{kambe_nmr_2013} \footnote{H. Amitsuka, \textit{Workshop on hidden order, superconductivity and magnetism in URu$_2$Si$_2$}, Leiden (2013).}.
Preliminary connections between the fingerprints of the HO transition have been made. The electronic structure of the HO is consistent with a periodicity given by the commensurate wave vector Q$_0$ \cite{oppeneer_electronic_2010} and part of the gapping of the incommensurate spin fluctuations was related to the loss of entropy at the HO transition \cite{wiebe_gapped_2007}. However, these relationships remain indirect. Additionally, the question of the symmetry of the novel excitations emerging from the HO state has not been addressed experimentally.

\par
In this letter using electronic Raman spectroscopy, we report clear Raman signatures of the HO state, i.e. a gap below $\sim55$~$\icm$(6.8~meV) and a sharp excitation deep inside the gap at 14~$\icm$(1.7~meV). Both signatures are observed only in the A$_{2g}$ symmetry, which indicates a direct intimate relationship between them. They emerge from a broad A$_{2g}$ quasi-elastic continuum which persists up to 300~K. Given the peculiarity of the A$_{2g}$ symmetry itself, our results give new and strong constraints on the nature of the HO state. We further demonstrate that the sharp Raman excitation tracks the resonance at \textbf{Q$_0$}=(0,0,1) detected by inelastic neutron scattering (INS) \cite{bourdarot_precise_2010} in the HO state as a function of temperature and magnetic field, indicating that both excitations have the same origin even if measured at different wave vector transfer. This brings a new and robust evidence for a Brillouin zone folding which places the $Z$ point on top of the $\Gamma$ point as expected in a transition between a body center tetragonal (bct) and a simple tetragonal (st) phase.

\par
Polarized Raman experiments have been carried out using a solid state laser emitting at 561~nm and a Jobin Yvon T64000 triple substractive grating spectrometer equipped with a nitrogen cooled CCD camera. Single crystals of URu$_2$Si$_2$ were grown by the Czochralski method using a tetra-arc furnace \cite{aoki_field_2010}. Three samples from the same batch with a residual resistivity ratio of $\sim50$ and freshly cleaved along the (ab) plane have been measured. Temperature and magnetic field dependencies have been performed in a closed-cycle $^{4}$He cryostat with sample in high vacuum and a $^{4}$He pumped cryostat with the sample in exchange gas, respectively \footnote{See Supplemental Material [http://link.aps.org/ supplemental/...], which includes Refs. \cite{hackl_gap_1983, maksimov_investigations_1992, mialitsin_raman_2010, aliev_anisotropy_1991}, for details about the laser heating estimation.}\nocite{hackl_gap_1983, maksimov_investigations_1992, mialitsin_raman_2010, aliev_anisotropy_1991}. By combining different incident and scattered light polarizations and sample geometry, we have extracted the A$_{1g}$, B$_{1g}$, B$_{2g}$ and A$_{2g}$ symmetries of the D$_{4h}$ point group (space group n$^\circ139$) \cite{hayes_scattering_2004}.

\begin{figure}[h]
\centering
\includegraphics[width=1\linewidth]{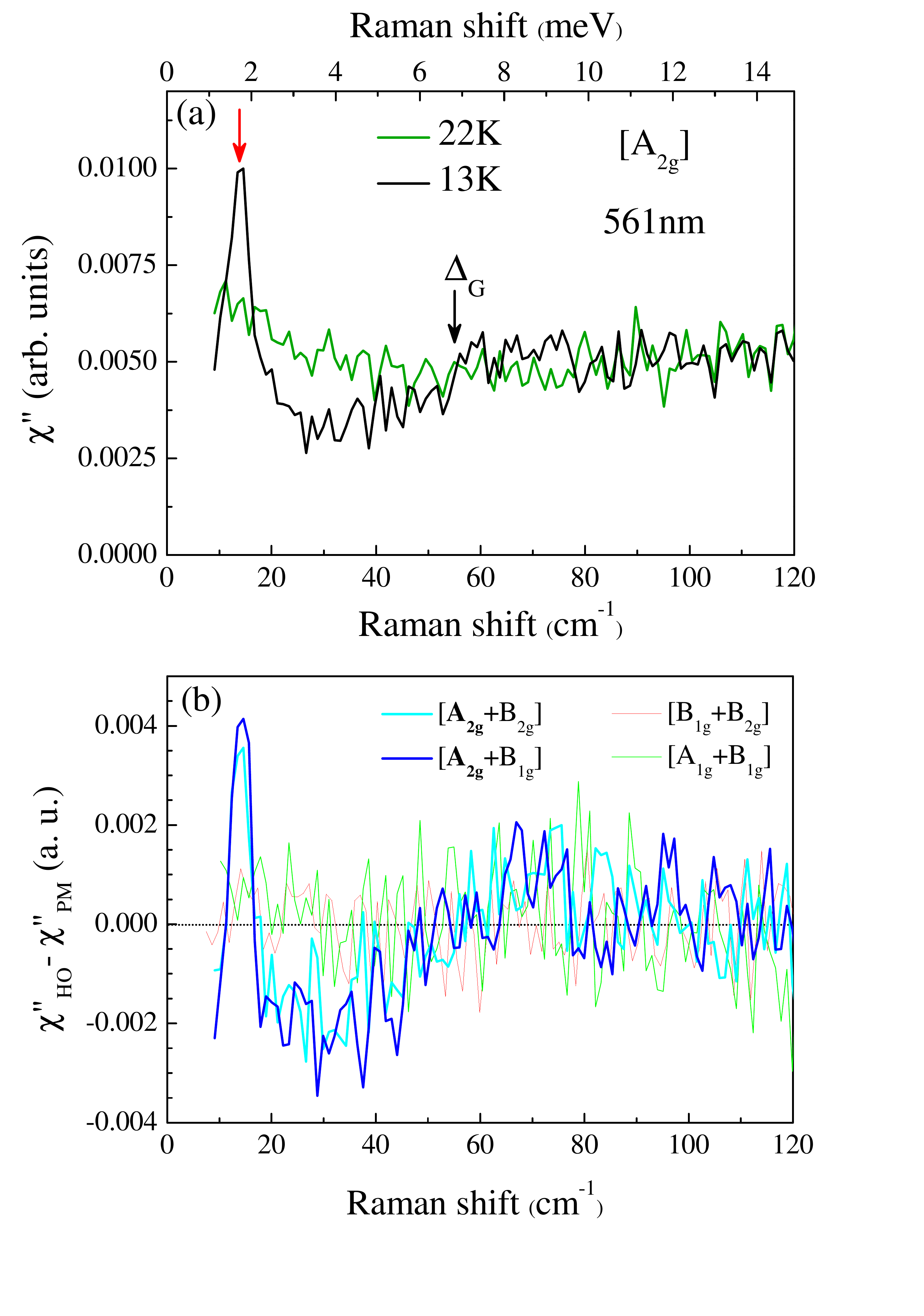}
\caption{(Color online) (a) Raman spectra of URu$_2$Si$_2$ in the pure A$_{2g}$ symmetry in the hidden order (HO) phase (13~K) and in the paramagnetic phase (22~K). In the HO, a sharp peak at 14~$\icm$, indicated by the red arrow, is superimposed on a gap below $\Delta_G\thicksim$~55~$\icm$. (b) Subtracted Raman responses in the hidden order phase and the paramagnetic phase for all probed symmetries. All spectra have been measured with the same laser power. Those containing the B$_{1g}$ response have been normalized to the intensity of the B$_{1g}$ phonon.}
\label{fig1}
\end{figure}

\par
Figure~\ref{fig1}(a) shows the low-frequency Raman spectra of URu$_2$Si$_2$ in the pure A$_{2g}$ symmetry~\footnote{See Supplemental Material [http://link.aps.org/ supplemental/...] for details about the extraction of the pure A$_{2g}$ symmetry.}. Upon entering the HO state, a gap opens below $\Delta_G\thicksim55\icm$ and a sharp excitation emerges at $\thicksim14\icm$ deep inside the gap. The subtraction of the Raman responses in the paramagnetic (PM) and HO state in various symmetries is reported Figure~\ref{fig1}(b). Within our accuracy, no feature can be detected in the other probed symmetries \footnote{See Supplemental Material [http://link.aps.org/ supplemental/...] for details about the measurements of the Raman response in other symmetries.}\footnote{We are aware of a parallel Raman study by the group of G. Blumberg with similar results \cite{kung_chirality_2014} interpreted in the context of the theory developed by Haule and Kotliar \cite{haule_arrested_2009}}\nocite{kung_chirality_2014}. The HO state is thus characterized by two Raman signatures of pure A$_{2g}$ symmetry. The A$_{2g}$ symmetry is intriguing. It is equivalent to the C$_{4h}$ subgroup of the D$_{4h}$ space group and transforms as $xy(x^2 -y^2)$ or $R_z$ \footnote{See Supplemental Material [http://link.aps.org/ supplemental/...], which includes Ref. \cite{harima_why_2010}, for details about the A$_{2g}$ symmetry.}\nocite{harima_why_2010}. This symmetry is usually associated with time reversal and/or chiral symmetry breaking excitations \cite{khveshchenko_raman_1994, liu_novel_1993, yoon_raman_2000, sulewski_observation_1991}. Among the large families of compounds belonging to the $D_{4h}$ space group, including the cuprates and the Fe-based superconductors, few measurements only have reported a sizeable A$_{2g}$ Raman response \cite{devereaux_inelastic_2007}.
\par
The energy of the gap is consistent with previous optical conductivity measurements~\cite{bonn_far-infrared_1988, guo_hybridization_2012, nagel_optical_2012, hall_observation_2012}. In addition, the gaps $\Delta_G$ extracted from the resistivity ($\sim$~56~$\icm$)~\cite{mcelfresh_effect_1987} and from heat capacity measurements ($\sim$~88~$\icm$)~\cite{maple_partially_1986, fisher_specific_1990} are in the same energy range than our findings. Scanning tunneling microscopy experiments also report a gap of $\sim\pm$4~meV, i.e. $\Delta_G~\simeq$~65~\icm \cite{aynajian_visualizing_2010, schmidt_imaging_2010}.

We note that Raman scattering is also sensitive to double excitations processes, such as double phonon or two-spin excitations with $\pm\textbf{Q}$ transferred wave-vectors. A double excitation process involving the resonance at $\pm$\textbf{Q$_1$} would be measured at $\sim$~75~$\icm$, in the energy range of the Raman gap. However, we can rule out this interpretation because the \textbf{Q$_1$} resonance strongly shifts to lower energy with increasing temperature. It has been reported to be inelastic above $T_0$ reaching $\sim$~2.5~meV at 20~K, i.e. 40~\icm for a double excitation \cite{broholm_magnetic_1991, bourdarot_neutron_2014} whereas the A$_{2g}$ Raman gap depletion vanishes at T$_0$ while its energy remains roughly constant up to T$_0$ \footnote{See Supplemental Material [http://link.aps.org/ supplemental/...] for details about the temperature dependence of the spectral weight of the A$_{2g}$ features.}. This comparison, as well as the observation of a similar gap by optical conductivity measurements, suggests that the depletion is not linked to the resonance at \textbf{Q$_1$} but to a gapped electron-hole excitations continuum as expected from a reconstruction of the Fermi surface inside the HO state \cite{santander-syro_fermi-surface_2009,yoshida_signature_2010,boariu_momentum-resolved_2013,elgazzar_hidden_2009}. We provide here a new information on this electron-hole excitations continuum, i.e. it has the pure A$_{2g}$ symmetry. From that, we surmise that the gap occurs in a continuum involving quasi-particles with strong spin-orbit character as described in \cite{oppeneer_spin_2011}.

\par

\begin{figure}[h!]
\centering
\includegraphics[width=1\linewidth]{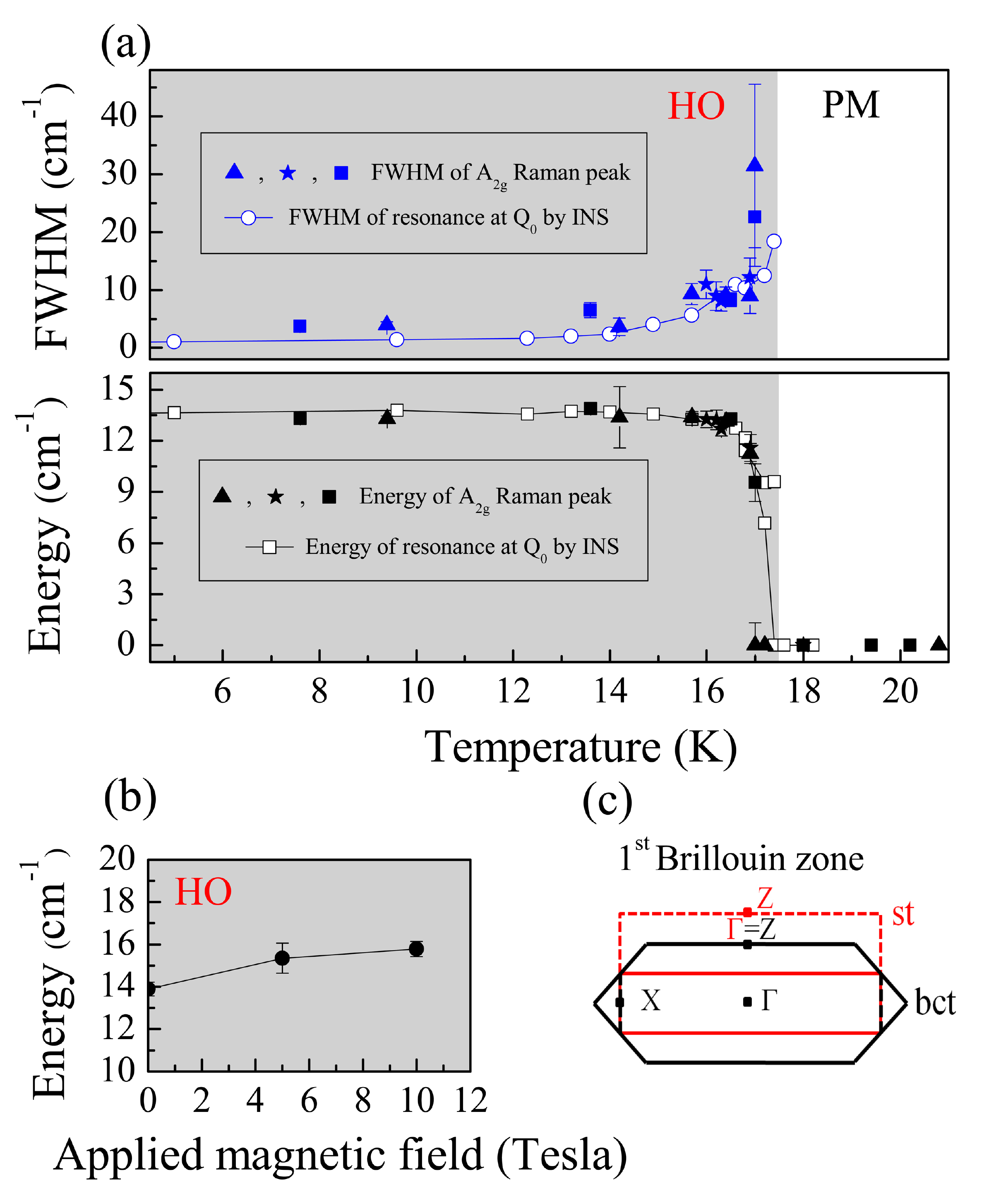}
\caption{(Color online) (a) Temperature dependence of the energy and full width at half maximum (FWHM) (after deconvolution from the resolution of the spectrometer (2~$\icm$)) of the sharp A$_{2g}$ Raman peak (full symbols) and of the resonance at \textbf{Q$_0$}=(0,0,1) measured by inelastic neutron scattering (INS) (open symbols) ~\cite{bourdarot_precise_2010}. The full symbols (stars, triangle, square) are extracted from various measurements on different samples from the same batch. The grey area corresponds to the temperature range of the HO phase. Both features are due to the same excitation probed by different techniques. The folding sketched in (c) would explain the observation of the same excitation at different \textbf{Q} wave vectors. (b) Magnetic field dependence of the energy of the sharp A$_{2g}$ Raman peak. The magnetic field is applied at 30$\pm5^\circ$ of the c axis. (c) Sketch of the first Brillouin zone in the paramagnetic state (body centered tetragonal, bct) and the folded one to the simple tetragonal (st) structure which folds the Z point to the $\Gamma$ point.}
\label{fig3}
\end{figure}

Below the gap, the A$_{2g}$ peak is sharp with a full width at half maximum (FWHM) of $\sim$~1$\icm$ at $\sim$~10~K, showing it is a long lived excitation. Figure~\ref{fig3}(a) presents the temperature dependence of its position and FWHM. Its energy and width are almost constant up to $\sim$~15~K. It abruptly drops to zero near T$_0$ with a temperature dependence stronger than expected for a mean field transition. It also broadens when approaching T$_0$. We also report the energy E$_0$ and FWHM of the neutron resonance at Q$_0$=(0,0,1), the up-to-now major signature of the HO phase~\cite{bourdarot_precise_2010, villaume_signature_2008}. The A$_{2g}$ Raman peak closely tracks the neutron resonance which strongly suggests that the same excitation is coupled to both probes. In addition, as shown figure~\ref{fig3}(b) the peak hardens slightly under magnetic field up to 10~T \footnote{See Supplemental Material  [http://link.aps.org/ supplemental/...], which includes Ref.  \cite{mentink_gap_1996}, for details about the magnetic field dependence of the Raman peak.}\nocite{mentink_gap_1996} which is qualitatively consistent with the magnetic field dependence of the neutron resonance \cite{bourdarot_evidence_2004}.
Raman spectroscopy probes the $\Gamma$ point, i.e. the total transferred wave vector $\textbf{Q}=0$. The feature observed at 1.7~meV by neutron scattering is measured at the $Z$ point, i.e. at Q$_0$=(0,0,1). Measuring the same excitation at the $\Gamma$ and $Z$ points can be explained by invoking a Brillouin zone folding along the c axis which occurs upon entering the HO state, as a bct to st transition would produce (Cf. Figure~\ref{fig3}(c)). The same conclusion was previously made from the comparison between the Fermi surfaces at ambient pressure and under pressure in the antiferromagnetic state \cite{hassinger_similarity_2010} as well as from the comparison between photoemission spectroscopy (ARPES) data at the $\Gamma$ and the $Z$ points \cite{yoshida_signature_2010, yoshida_translational_2013, boariu_momentum-resolved_2013, meng_imaging_2013, bareille_momentum-resolved_2014}. The conclusion drawn here is robust as it results from measurements at zero magnetic field and zero applied pressure. Most of all it is based on the observation of a major signature of the HO state at different \textbf{Q} vectors.

\par

\begin{figure}[h!]
\centering
\includegraphics[width=1\linewidth]{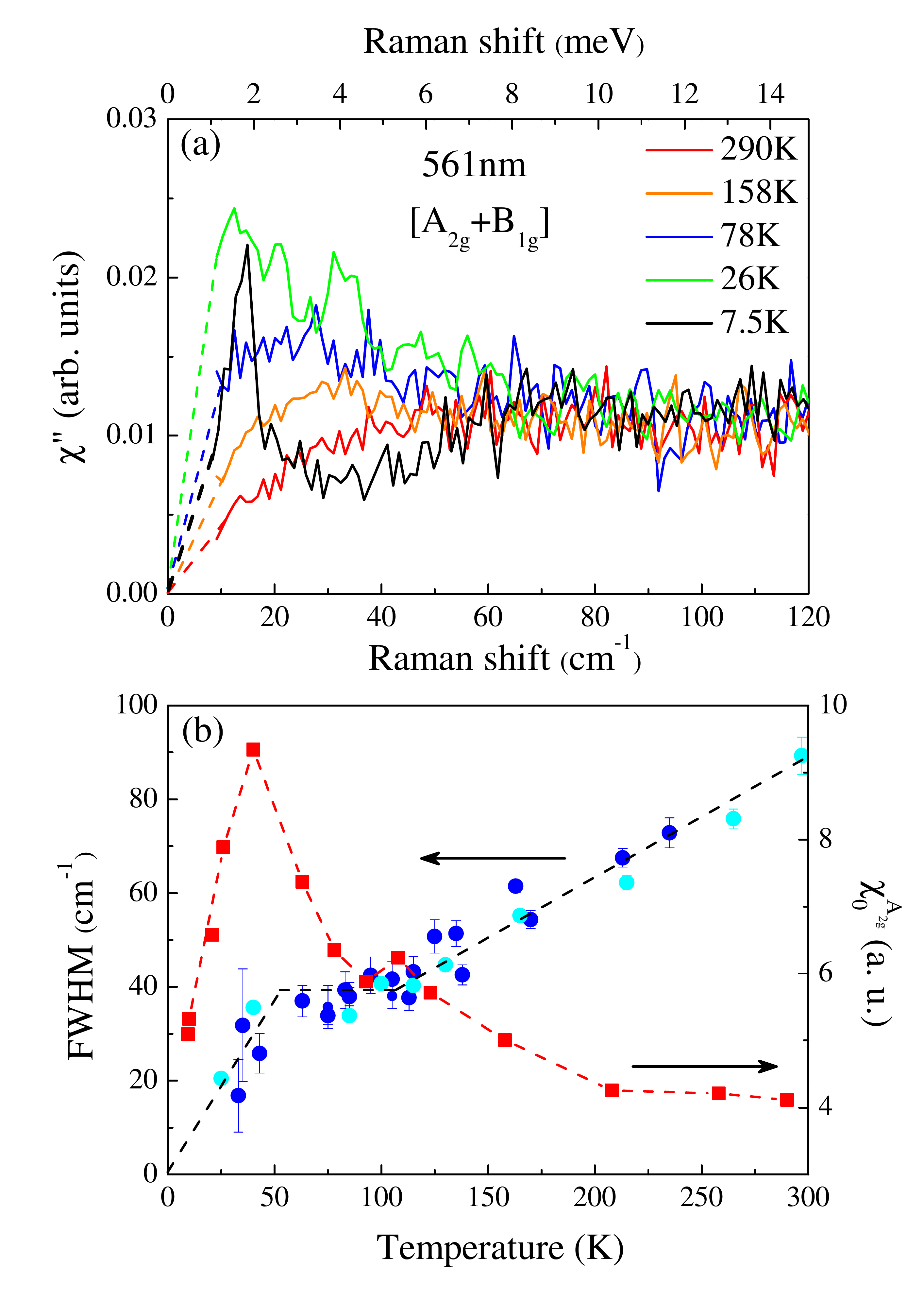}
\caption{(Color online) (a) Raman susceptibility in the A$_{2g}$+B$_{1g}$ symmetry from 7.5~K to 300~K. (b) Left axis: Full width of the quasi-elastic response (Cf. text) (blue filled circles) versus temperature. Right axis: Raman static susceptibility $\chi^{A_{2g}}_0$ (red filled squares) versus temperature (Cf. text). Dashed lines are guide for the eyes.}
\label{fig2}
\end{figure}

Interestingly, the finite Raman response in the A$_{2g}$ symmetry is not limited to the HO state. Indeed, as reported by Cooper et al. \cite{cooper_raman_1986}, already at 300~K, URu$_2$Si$_2$ exhibits a A$_{2g}$ quasi-elastic peak (QEP) with a overdamped Lorentzian line shape. As shown Figure~\ref{fig2}(a), the quasi-elastic contribution sharpens with decreasing temperature before collapsing in the HO state. The FWHM, calculated with a simple relaxation model \cite{cooper_raman_1986}, is reported in Figure~\ref{fig2}(b). After a Korringa-like linear temperature dependence down to $\sim100$ K, it exhibits a plateau-like behavior between $\sim100$ K and $\sim50$ K before decreasing three times faster down to 20~K. The plateau is thus limited to the Kondo regime with an increase of the lifetime of the A$_{2g}$ excitations most probably below the Kondo coherence temperature. Via the Kramers-Kronig relation \cite{gallais_observation_2013}, the A$_{2g}$ static susceptibility $\chi^{A_{2g}}_0$ can be extracted as $\int \! \chi_{A_{2g}}''(\omega)/\omega \, \mathrm{d}\omega$ with integration spanning from 8 to 100~\icm, above which all spectra are on top of each other. As shown in Figure~\ref{fig2}(b), the $A_{2g}$ static susceptibility exhibits a temperature dependence very reminiscent to the dc magnetic susceptibility along the c axis \cite{palstra_magnetic_1986}, suggesting a link between the A$_{2g}$ degree of freedom and the magnetic susceptibility. The temperature dependence of this last one has been tentatively explained considering various crystalline electric field (CEF) schemes \cite{santini_crystal_1994}.
A similar Raman quasi-elastic response has already been discussed in the context 4f and 5f system, where it was attributed to either spin fluctuations or localized CEF excitations like in UBe$_{13}$ \cite{cooper_polarized_1988}. In URu$_2$Si$_2$, a tempting simple interpretation of the A$_{2g}$ QEP would be to consider a CEF excitation between very broad (and partially delocalized) levels. Indeed, simple local CEF excitations on the U atoms can have a A$_{2g}$ symmetry, with different ground state and with an even or odd number of localized electrons \footnote{See Supplemental Material [http://link.aps.org/ supplemental/...], which includes Ref. \cite{koster_properties_1963}, for details about the CEF excitations selection rules.}\nocite{koster_properties_1963}.

Following this interpretation, a global and simple scenario, reminiscent of the results obtained across the metal-insulator transition of the skutterudite PrRu$_4$P$_{12}$ \cite{iwasa_evolution_2005}, can be given. The A$_{2g}$ QEP as well as the sharp A$_{2g}$ peak in the HO state are due to CEF excitations while the gap $\Delta_G$ is associated to the gapped itinerant electron-hole continuum. Because CEF excitations are coupled to itinerant carriers, the QEP is strongly damped and quasi-elastic above T$_0$. At the HO transition, A$_{2g}$ decaying channels are quenched at low energy due to the opening of the A$_{2g}$ gap in the electronic continuum and consequently, the CEF excitation becomes long-lived. By the same process the CEF excitation becomes gapped (inelastic) because of the associated loss of hybridization with delocalized quasiparticle continuum in the HO state. In this picture, the dual character of the phase transition in URu$_2$Si$_2$ appears naturally, with the gap $\Delta_G$ observed in the Raman continuum, directly linked to the itinerant nature of the 5f electrons and the CEF excitation observed as the Raman sharp peak, associated with their local character. Within this scenario, a close relationship between the signatures of the itinerant and localized character of the HO state can be made, through the similarity of their A$_{2g}$ symmetry.

Now, we discuss alternative explanations for each signature, the QEP and the sharp peak, independently. We inferred that the A$_{2g}$ gap is a gapped electron-hole excitation continuum. As already noted above, the opening of this gap will help any excitation inside the gap to become long-lived and sharp. In this context, the peculiar A$_{2g}$ symmetry may emerge from local current loop excitations, which could give rise to the A$_{2g}$ peak in the HO state. Indeed similar anomalous orbital motion of charge carriers have been shown to have the A$_{2g}$ symmetry \cite{khveshchenko_raman_1994} based on a Raman experiment in the insulating cuprates \cite{liu_novel_1993}. This hypothesis has also been recently brought up for URu$_2$Si$_2$ \cite{mydosh_hidden_2014,oppeneer_spin_2011}.
As for the QEP observed above T$_0$, in a pure itinerant picture perspective, a Drude-like Raman response of electron-hole excitations may explain it. We would need then to account for the strong spectral weight in the A$_{2g}$ symmetry. This would certainly require to go beyond the effective mass approximation \cite{devereaux_inelastic_2007} by taking into account the resonant contributions for the Raman vertex calculation \cite{shvaika_electronic_2005,khveshchenko_raman_1994} \footnote{Indeed, in the effective mass approximation, the electron-hole excitations continuum is not A$_{2g}$ active except by considering an unlikely electronic band structure on which $\frac{\partial^2 E}{\partial k_x \partial k_y}\neq\frac{\partial^2 E}{\partial k_y \partial k_x}$ in some k-space region of the Fermi surface.}.



\par
In conclusion, we have reported two Raman features of pure A$_{2g}$ symmetry, a sharp peak at 14~\icm and a low energy gap at $\sim$~55~\icm, as signatures of the hidden order state in URu$_2$Si$_2$. Additionally, by performing accurate temperature and magnetic field dependencies of the sharp Raman peak, we have shown that the sharp peak matches the neutron resonance at \textbf{Q$_0$}. This brings new and robust evidence of the Brillouin zone folding along c axis upon entering the HO state, consistent with a switch from body center tetragonal to simple tetragonal. Theoretical investigations, most probably accounting for both the local and the itinerant character of the quasiparticle in URu$_2$Si$_2$, are necessary to reach a global scenario of these Raman signatures of peculiar A$_{2g}$ symmetry and to conclude about the associated HO order parameter.

\begin{acknowledgments}

This work was supported by the Labex SEAM (Grant No. ANR-11-IDEX-0005-02) and by the french Agence Nationale de la Recherche (ANR PRINCESS).
We thanks F. Bourdarot, S. Burdin, C. Pepin, G. Knebel, I. Paul, D. Khveshchenko, G. Blumberg, G. Kotliar, H. Harima and P. Oppeneer for very fruitful discussions.

\end{acknowledgments}

\bibliographystyle{apsrev4-1}
\bibliography{biblio}

\end{document}


\title{Supplemental Material : Symmetry of the Excitations in the Hidden Order State of URu$_2$Si$_2$}
\author{J. Buhot}
\author{M.A. M\'easson}
\author{Y. Gallais}
\author{M. Cazayous}
\author{A. Sacuto}
\affiliation{Laboratoire Mat\'eriaux et Ph\'enom\`enes Quantiques, UMR 7162 CNRS, Universit\'e Paris Diderot, B$\hat{a}$t. Condorcet 75205 Paris Cedex 13, France}
\author{G. Lapertot}
\author{D. Aoki}
\affiliation{Univ. Grenoble Alpes, INAC-SPSMS, F-38000 Grenoble, France}
\affiliation{CEA, INAC-SPSMS, F-38000 Grenoble, France}

\date{\today}

\maketitle

\section{Experimental Details}

\subsection{Laser heating estimation}

The focused laser light used for Raman experiment induces a local heating on the crystal surface. To study the hidden order of URu$_2$Si$_2$, we aimed to reach with certainty the HO state below T$_{0}$=17.5~K and to track accurately the temperature dependence of various features. Therefore, we used a low laser power, and we took special care to estimate the laser heating. For each experiment, the laser heating was estimated for a given bath temperature by using the method reported in \cite{hackl_gap_1983}. Then, the temperature dependence of this laser heating was determined, as quoted in~\cite{maksimov_investigations_1992, mialitsin_raman_2010}, by considering the temperature dependence of the thermal conductivity of URu$_2$Si$_2$ along the c axis \cite{behnia_thermal_2005, aliev_anisotropy_1991}. For a spot light of $\sim 50\mu m$ of diameter focused on samples in vacuum, the laser heating is $\sim2$ K/mW for a bath temperature of 4~K and $\sim1$ K/mW for a bath temperature of 30~K. The reported temperatures are estimated to have an accuracy of $\pm 1~K$.

\subsection{Extraction of the pure A$_{2g}$ symmetry}

For the D$_{4h}$ space group, there is no configuration of incident and scattered polarizations of light which provides a pure A$_{2g}$ Raman response. Nevertheless, we were able to extract a pure A$_{2g}$ signal using simple operations between Raman responses of mixed symmetries. We have added the following contributions:
\begin{equation}
   [A_{2g}+B_{1g}]+[A_{2g}+B_{2g}]=[2A_{2g}+B_{1g}+B_{2g}]
   \label{eq1}
   \end{equation}
Thanks to the use of circular polarizations, we have access to the symmetries:
\begin{equation}
   [B_{1g}+B_{2g}]
   \label{eq2}
   \end{equation}
Thus, by removing the contribution (\ref{eq2}) from the contribution (\ref{eq1}), we obtained a pure A$_{2g}$ signal.

\section{Symmetry dependence of the Raman signal}

\begin{figure*}[htbp]
\centering
\includegraphics[width=0.9\linewidth]{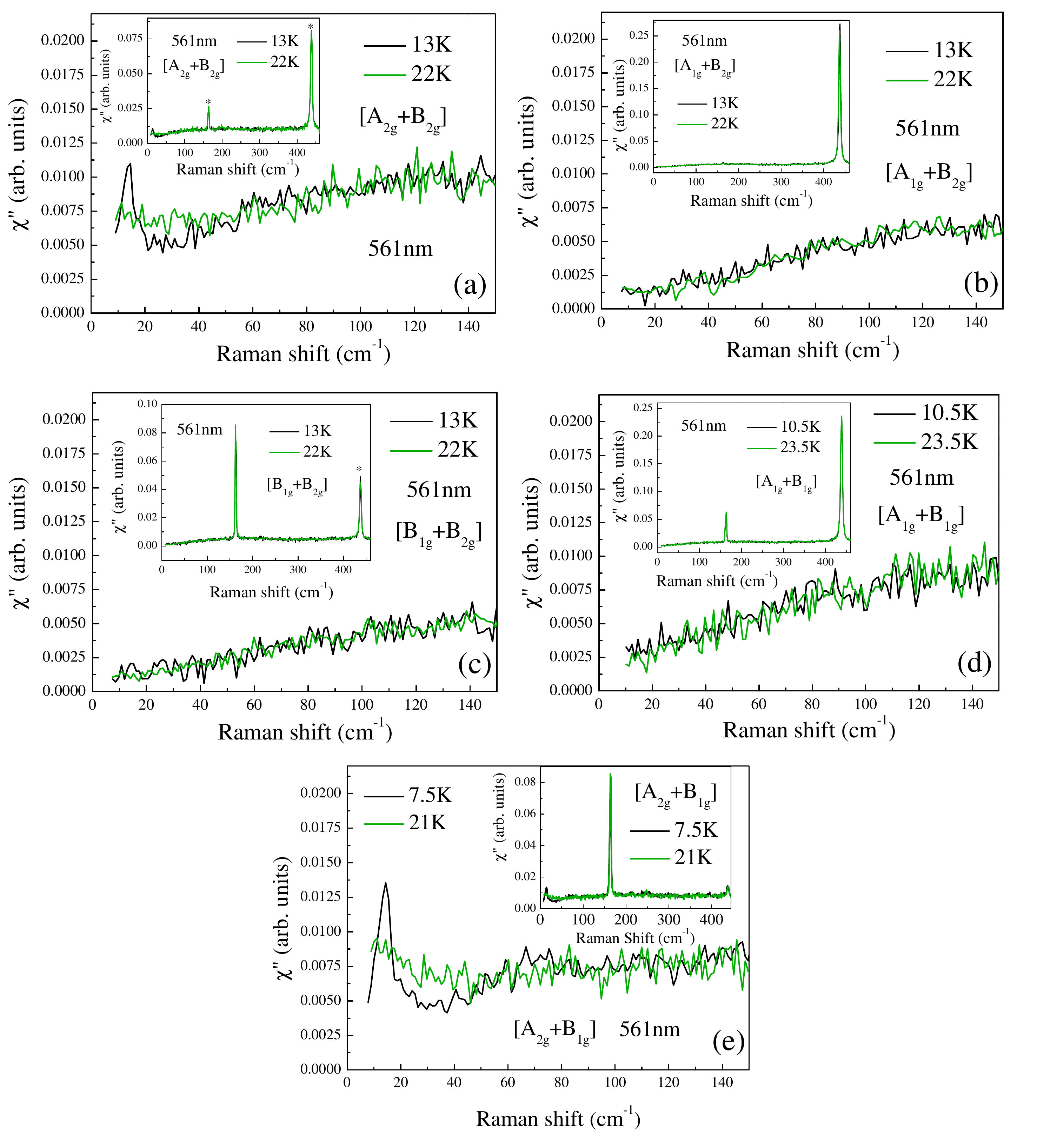}
\caption{(Color online) Raman spectra of URu$_2$Si$_2$ in the HO phase (black line) and in the PM phase (green line) in the A$_{2g}$+B$_{2g}$ (a), A$_{1g}$+B$_{2g}$ (b), B$_{1g}$+B$_{2g}$ (c),  A$_{1g}$+B$_{1g}$ (d) and A$_{2g}$+B$_{1g}$ symmetries (e). All spectra have been measured with the same laser power and the same optical set-up. Large energy range with the same intensity scale is shown in the insets for each symmetries. Leakage of phonon modes are marked by a black star.}
\label{fig5}
\end{figure*}

Figure~\ref{fig5} shows the Raman spectra of URu$_2$Si$_2$ in all symmetries, i.e. mixtures of A$_{1g}$, B$_{1g}$, A$_{2g}$ and B$_{2g}$, when the polarizations of light are within the (a,b) plane. Only the E$_g$ symmetry cannot be probed in this configuration. The measurements were done in the hidden order state (HO) (10.5~K and 13~K) and in the paramagnetic (PM) state (22~K and 23~K). In the HO state, the [A$_{2g}$+B$_{2g}$] and the [A$_{2g}$+B$_{1g}$] symmetries responses exhibit a gap opening below 55~\icm and a sharp excitation at $\thicksim14~\icm$ deep inside the gap (See Figure~\ref{fig5} (a) and (e)). These features are not observed in the other symmetries, \textit{i.e.} [A$_{1g}$+B$_{2g}$], [B$_{1g}$+B$_{2g}$] and [A$_{1g}$+B$_{1g}$] contributions (See Figure~\ref{fig5} (b), (c), (d), respectively). Therefore, we concluded that the two signatures of the HO state have A$_{2g}$ symmetry~\footnote{We are aware of parallel study by G. Blumberg et al. who reports peaks in the A$_{2g}$ and A$_{1g}$ symmetry. Within our experimental accuracy (noise $\sim$ 1/5 of the A$_{2g}$ peak), we cannot resolve any peak in the A$_{1g}$ symmetry.}.  For comparison, the spectra were reported without any normalization, and they were measured with the same laser power and the same optical set-up (except for the insertion of polarizers). We also checked that in the [B$_{1g}$+B$_{2g}$] symmetry the Raman response is the same at 35~K as at 13~K and at 22~K. For comparison, we presented a large energy scale in the insets of the Figure~\ref{fig5}, in order to show the phonon mode intensities. The A$_{1g}$ and B$_{1g}$ phonon modes are visible here. For some symmetries, a leakage of the phonon modes due to the crystal misalignment is observed (See insets of Figure~\ref{fig5} (a) and (c)). Within our accuracy, there is no signature of charge nematic fluctuations \cite{gallais_observation_2013}, neither in B$_{1g}$ nor in B$_{2g}$ symmetry. We cannot provide any additional evidence for the fourfold symmetry breaking reported by torque \cite{okazaki_rotational_2011} and by cyclotron resonance experiments \cite{tonegawa_cyclotron_2012}.

\section{Temperature dependence of the Raman signatures of the Hidden order state}

\begin{figure}[htbp]
\centering
\includegraphics[width=1\linewidth]{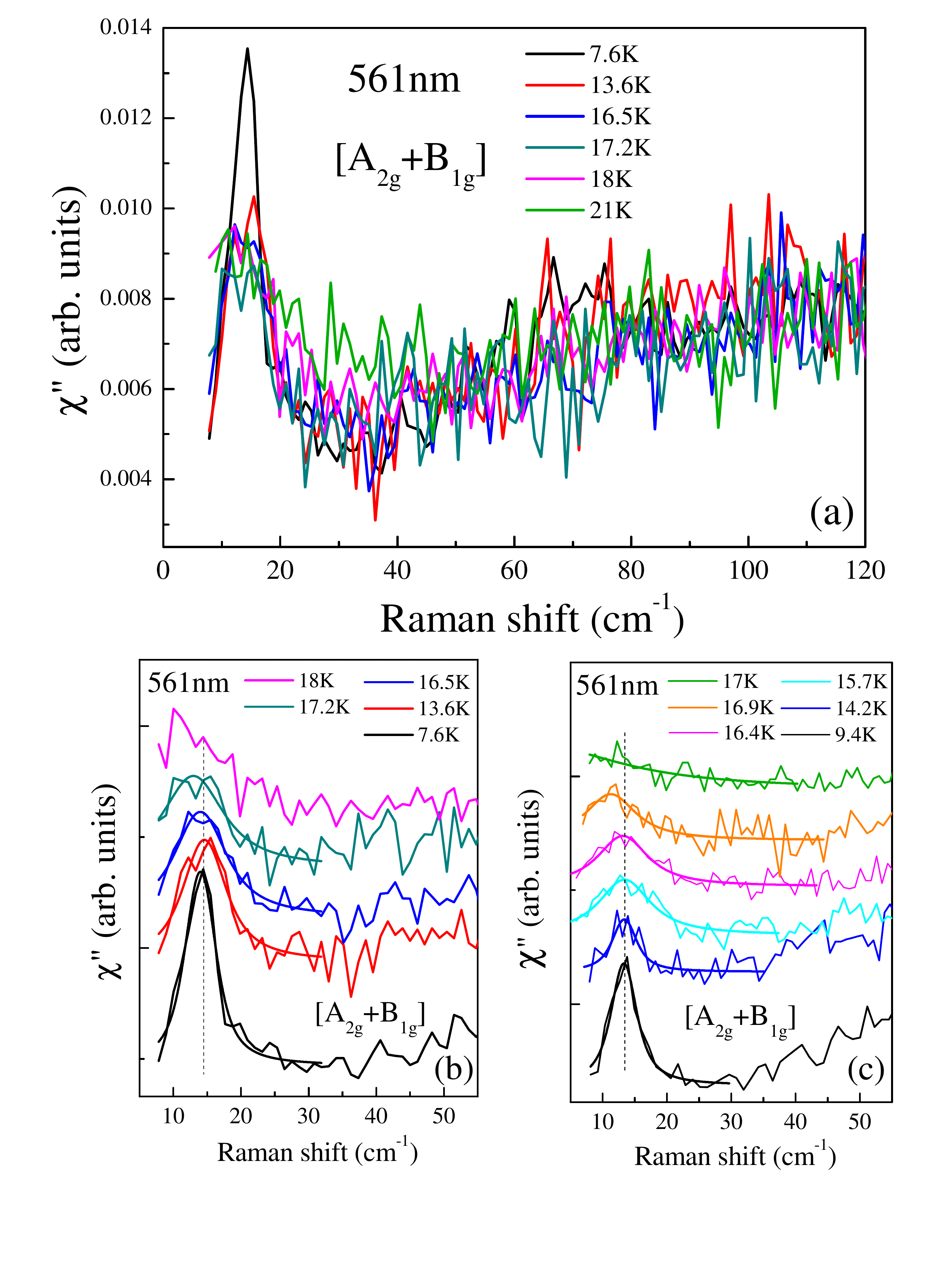}
\caption{(Color online) Raman spectra of URu$_2$Si$_2$ in the A$_{2g}$+B$_{1g}$ symmetry at several temperatures across the hidden order transition. All the temperatures are estimated at $\pm1~K$. (a) and (b) Data extracted from an experiment with sample in vacuum. (c) Data extracted from an experiment with sample in exchange gas. In (b) and (c), the spectra focused on the A$_{2g}$ peaks are shifted for clarity. The lines are fits done with a Voigt function taking into account the resolution of the spectrometer.}
\label{fig7}
\end{figure}

Figure~\ref{fig7} shows the [A$_{2g}$+B$_{1g}$] Raman response of URu$_2$Si$_2$ at various temperatures below 25~K. As shown in panel (a), the A$_{2g}$ Raman gap closes abruptly near T$_0$, with a filling of the gap starting above 17.2~K. The energy of the gap remains roughly constant until it closes completely. Figure~\ref{fig7}(b) and (c) present the spectra focused on the A$_{2g}$ peak with data measured on a sample in vacuum (also presented Figure~\ref{fig7}(a)), and with data measured during another experiment with sample in exchange gas, respectively. The A$_{2g}$ peak continuously loses spectral weight and already starts to soften while the gap is still fully open, as evidenced in Figure~\ref{fig7}(b). Above about 16.5~K, the A$_{2g}$ peak abruptly softens and broadens. The peaks are fitted with a Voigt profile which takes into account the resolution of the spectrometer. The values of the position and FWHM extracted from these fits are reported in Figure~2 of the Letter as full squares and triangles for sample in vacuum and in exchange gas, respectively.

\begin{figure}[htbp]
\centering
\includegraphics[width=1\linewidth]{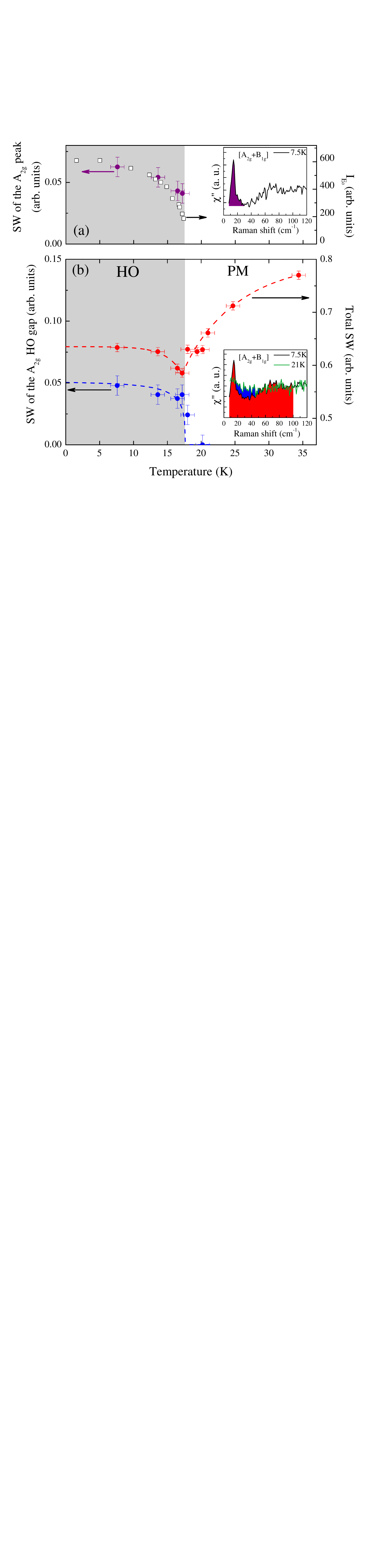}
\caption{(Color online) (a) Spectral weight of the A$_{2g}$ sharp Raman peak and integrated imaginary part of the neutron dynamical spin susceptibility at \textbf{Q$_0$}, I$_{E_0}$ \cite{bourdarot_precise_2010} versus temperature. Inset: illustration of the calculation of the spectral weight of the Raman peak. (b) Spectral weight of the HO gap calculated just above the A$_{2g}$ peak and total spectral weight up to 100~\icm. The A$_{2g}$ spectral weight is not conserved. The gap closes abruptly near T$_0$. Inset: illustration of the calculation of the spectral weights.}
\label{fig6}
\end{figure}

We evaluated the spectral weights of the sharp peak and the gap depletion in the HO state and the total spectral weight by integrating $\chi''(\omega)$. The spectral weight of the sharp peak smoothly decreases with decreasing temperature (See figure \ref{fig6}(a)). The integration of the magnetic resonance at \textbf{Q$_0$} measured by INS is also reported figure \ref{fig6}(a). The T-dependence of the two spectral weights roughly follows the same trend. Discussing a precise temperature dependence (BCS type) of the Raman peak is beyond our accuracy. As clearly shown in figure \ref{fig6}(b), the total spectral weight is neither conserved in the HO state between the gap and the sharp peak nor between the HO state and the PM state. We conclude that no clear spectral weight transfer exists between the gap and the peak. We observed the closing of the gap at 18~K$\pm1$K, while the peak becomes broader and quasi-elastic.

\section{Magnetic field dependence of the HO signatures}

\begin{figure}[htbp]
\centering
\includegraphics[width=0.7\linewidth]{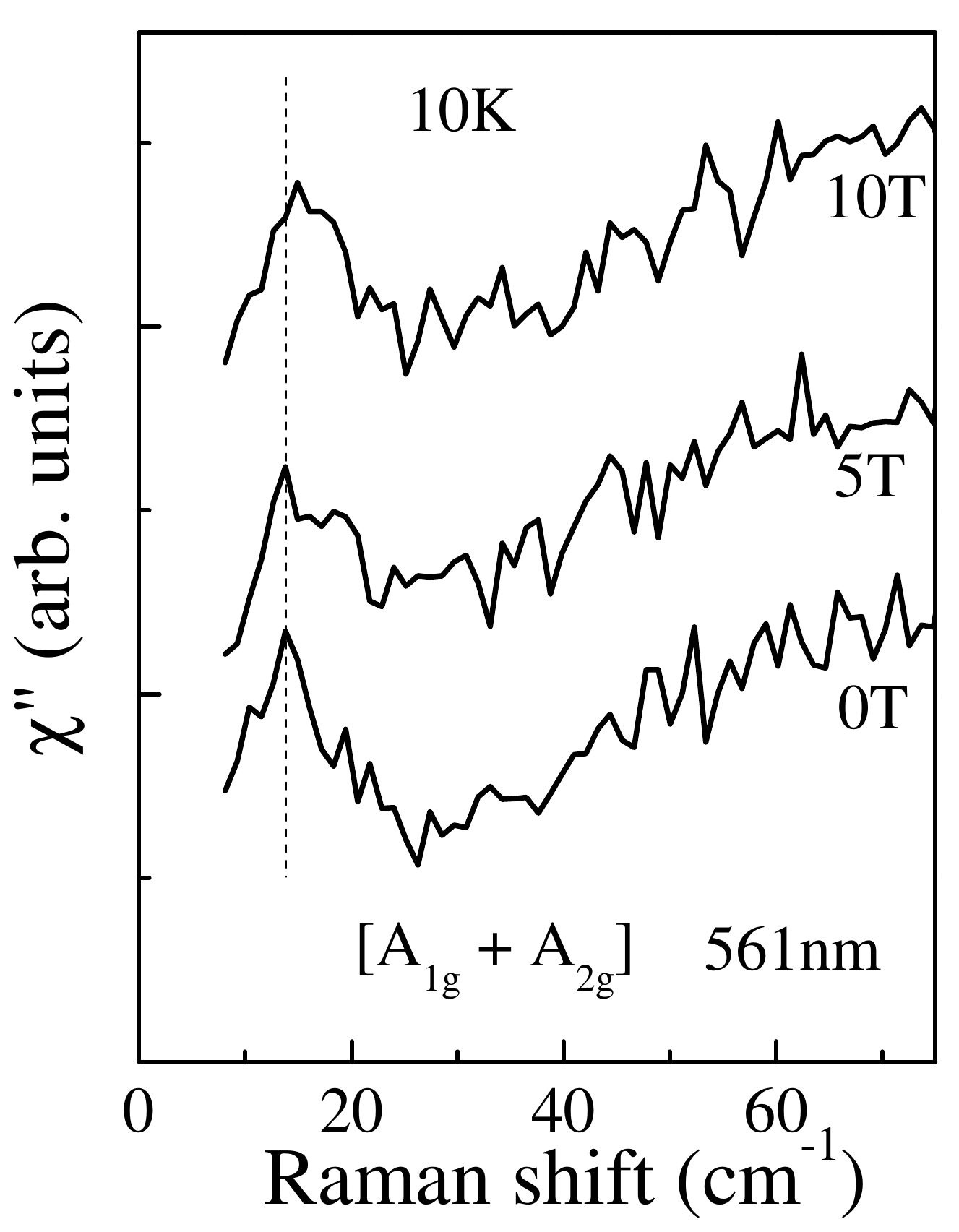}
\caption{(Color online) (a) Raman spectra of URu$_2$Si$_2$ in the A$_{2g}$+A$_{1g}$ symmetry in the HO state (10~K) at 0, 5 and 10~T. Data have been shifted for more clarity and the same background has been subtracted from the raw data. The A$_{2g}$ sharp peak slightly hardens and the A$_{2g}$ gap is roughly stable with increasing magnetic field.}
\label{fig4}
\end{figure}
The magnetic field dependence of the Raman response has been performed in a $^{4}$He pumped cryostat with the sample in exchange gas using circular polarizations. As shown Figure~\ref{fig4}, the peak hardens slightly under magnetic field up to 10~T which is qualitatively consistent with the magnetic field dependence of the neutron resonance E$_0$ \cite{bourdarot_evidence_2004}. The peak does not broaden under magnetic field. The A$_{2g}$ gap remains open under magnetic field up to 10~T without any significative shift in energy, as expected from the resistivity measurements under magnetic field \cite{mentink_gap_1996}.

\section{A$_{2g}$ symmetry}

\begin{figure}[htbp]
\centering
\includegraphics[width=1\linewidth]{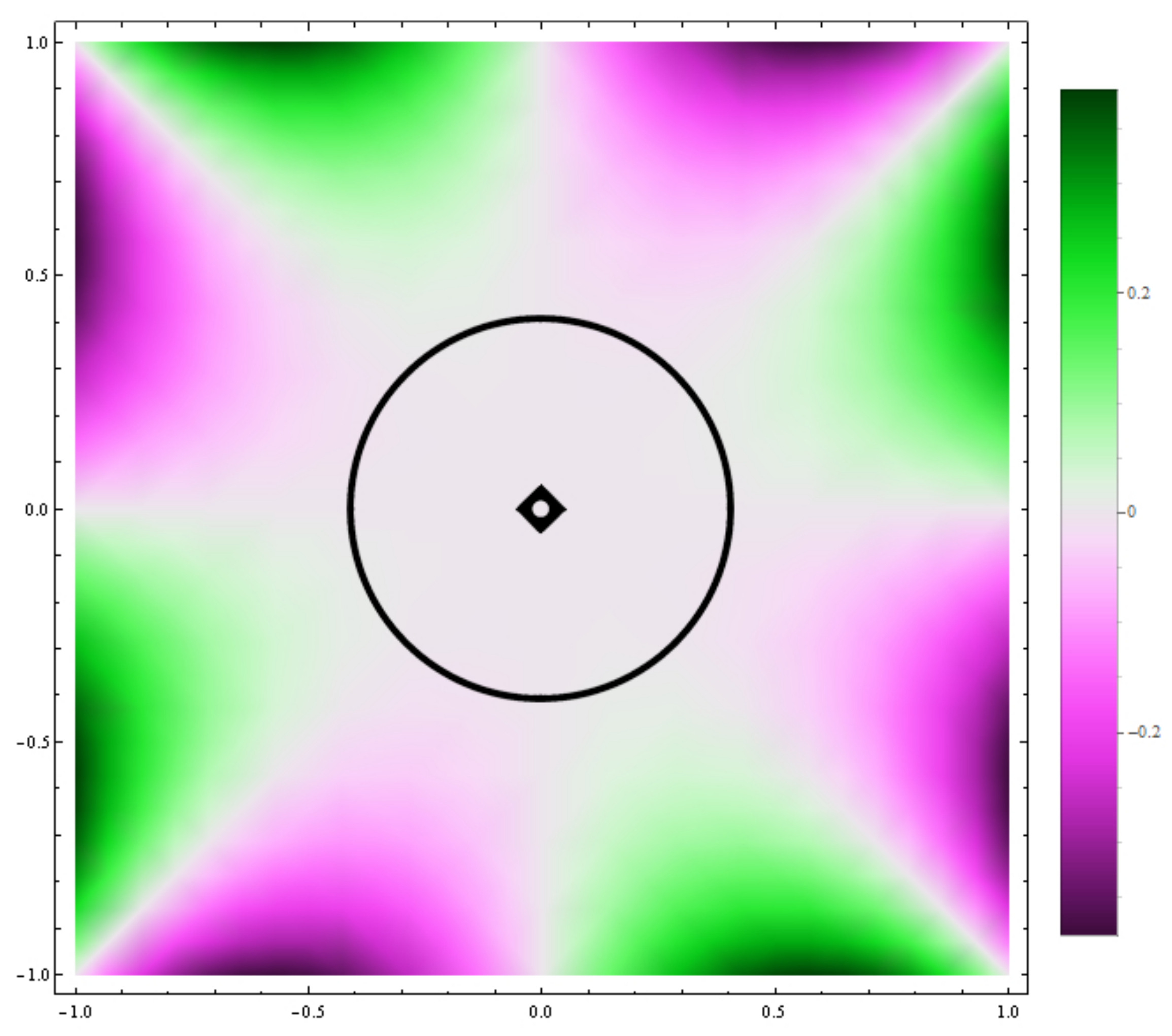}
\caption{(Color online) Representation of the A$_{2g}$ function $xy(x^2-y^2)$. Middle: Stereographic projection of the A$_{2g}$ symmetries (equivalent to the C$_{4h}$ subgroup of the D$_{4h}$ group).}
\label{fig8}
\end{figure}

Two distinct signatures of the HO state, namely the gap and the sharp peak, are reported to have a well-defined symmetry A$_{2g}$.
The A$_{2g}$ transforms like the function $ixy(x^2-y^2)$ (See Figure~\ref{fig8}) or $R_z$ (axial vector along z axis). It respects the 4-fold symmetry $C_4$ and the inversion center symmetry $I$. It lacks the two-fold axis symmetries C'$_2$(axis x and y) and C''$_2$ (axis (110)) and the reflections $\sigma_v$ (planes xz and yz) and $\sigma_d$ (planes$\bot(110)$ and $\bot(1\bar{1}0)$). Stereographic projection of the elementary A$_{2g}$ symmetries is shown Figure~\ref{fig8}. The A$_{2g}$ symmetry is equivalent to the C$_{4h}$ subgroup of the D$_{4h}$ group.

We emphasize that the order parameter of the HO state does not {\it necessarily} have the same A$_{2g}$ symmetry as the Raman signatures.

Now, we discuss the Raman selection rules in the space group of URu$_2$Si$_2$, i.e. n$^\circ$ 139, and in the 4 space groups selected by Harima et al. \cite{harima_why_2010}. All the space groups selected by Harima et al \cite{harima_why_2010} (n$^\circ$ 126, 128, 134, 136) as possible candidate for the space group of the HO state have the point symmetry D$_{4h}$. Then, the selection rules for the phonons are the same inside the HO state than in the PM state (group n$^\circ$ 139). However, the local point symmetry at the Uranium site depends on the space group as shown Table~\ref{tab1}. This will change the irreducible representations and then the selection rules for any local excitations based on the Uranium sites. Table~\ref{tab1} gives the mapping of the representations (of the PM state) into the representations of the new local point group (of the HO state). For the space groups n$^\circ$ 126 and 134, the mapping of the irreducible representations is equivalent to the one in the space group n$^\circ$ 139. Thus the Raman matrices and the selection rules are the same, including the ones for crystalline electric field (CEF) excitations. However for the space groups n$^\circ$ 128 and 136, the irreducible representations are different and the selection rules for local excitations on the Uranium sites have to be revisited.

\begin{table*}[htbp]
\centering
\renewcommand{\arraystretch}{1.8}
\begin{tabular}{|>{\centering}m{2cm}|c|c|c|c|c|}
\hline
Space group & n$^\circ$139 (D$_{4h}$)& n$^\circ$126 (D$_{4h}$) & n$^\circ$128 (D$_{4h}$) & n$^\circ$134 (D$_{4h}$) & n$^\circ$136 (D$_{4h}$) \\
\hline
Local Point symmetry at U & D$_{4h}$ & D$_{4}$ & C$_{4h}$ & D$_{2d}$ & D$_{2h}$ \\
\hline
\multirow{5}{2cm}{Irreductible Representations (simple)} &  $\Gamma_1^+$ (A$_{1g}$) & $\Gamma_1$ (A$_{1}$) &$\Gamma_1^+$ (A$_{g}$) & $\Gamma_1$ (A$_{1}$) &$\Gamma_1^+$ (A$_{g}$) \\
& $\Gamma_2^+$ (A$_{2g}$)  & $\Gamma_2$ (A$_{2}$) &$\Gamma_1^+$ (A$_{g}$) &$\Gamma_2$ (A$_{2}$) & $\Gamma_3^+$ (B$_{1g}$)\\
& $\Gamma_3^+$ (B$_{1g}$)  & $\Gamma_3$ (B$_{1}$) &$\Gamma_2^+$ (B$_{g}$) &$\Gamma_3$ (B$_{1}$) & $\Gamma_1^+$ (A$_{g}$)\\
& $\Gamma_4^+$ (B$_{2g}$)  &  $\Gamma_4$ (B$_{2}$) &$\Gamma_2^+$ (B$_{g}$) &$\Gamma_4$ (B$_{2}$) & $\Gamma_3^+$ (B$_{1g}$)\\
& $\Gamma_5^+$ (E$_{g}$)   & $\Gamma_5$ (E)      &$\Gamma_3^+$ ($^1$E$_g$) $\oplus$ $\Gamma_4^+$ ($^2$E$_g$)  & $\Gamma_5$ (E) & $\Gamma_2^+$ (B$_{2g}$) $\oplus$ $\Gamma_4^+$ (B$_{3g}$)\\
\hline
\multirow{2}{2cm}{Irreductible Representations (double)} & $\Gamma_6^+$ & $\Gamma_6$ & $\Gamma_5^+$ $\oplus$ $\Gamma_6^+$ & $\Gamma_6$ & $\Gamma_5^+$ \\
& $\Gamma_7^+$ &  $\Gamma_7$  & $\Gamma_7^+$ $\oplus$ $\Gamma_8^+$ & $\Gamma_7$ & $\Gamma_5^+$\\
\hline
\end{tabular}
\caption{Mapping of the irreducible representations of the point group D$_{4h}$ (space group of URu$_2$Si$_2$ and local point symmetry at the Uranium site in URu$_2$Si$_2$) into the irreducible representations of the new local point groups upon reduction of symmetry. The new local point groups define the symmetry at the Uranium site associated with the 4 space groups selected by Harima et al.\cite{harima_why_2010} as candidate for the HO state. Are shown the irreducible representations in the simple and double representation. The local point symmetry is relevant for the selection rules of localized excitations on the Uranium site such as CEF excitations.}
\label{tab1}
\end{table*}

Next, assuming that the A$_{2g}$ peak is due to a local excitation on the Uranium site such as a CEF excitation, if the HO state is described by the space groups n$^\circ$ 136, a A$_{2g}$ Raman signal would be accompanied by a signal in the B$_{2g}$ symmetry (note here that the symmetries we used are the ones of the D$_{4h}$ point group). For the space group n$^\circ$ 128 we would expect a non zero Raman signal in the A$_{1g}$ symmetry together with the A$_{2g}$ one \cite{kung_chirality_2014}. The relative intensity of these signals depends on the Raman matrices and cannot be assessed by symmetry arguments alone. For the space groups n$^\circ$ 126 and 134, the Raman response in the A$_{2g}$ channel is independent of the ones in all others symmetries, i.e. we do not expect, a priori, a signal in another symmetry.

We discuss now the selection rules for particular excitations, i.e. the CEF excitations. Table~\ref{tab2} provides all the CEF excitations that give rise to a signal in the A$_{2g}$ channel. In both cases, with an even (simple representation) or odd (double representation) number of localized electrons, some CEF excitations are active in the A$_{2g}$ symmetry. These results are extracted from the multiplication tables of the irreducible representations \cite{koster_properties_1963} and are independent on the respective positions in energy of the CEF levels. Some of the CEF transitions reported in table~\ref{tab2} are also active in few additional symmetries (see \cite{koster_properties_1963}).

\begin{table*}[htbp]
\centering
\renewcommand{\arraystretch}{1.8}
\begin{tabular}{|>{\centering}m{3cm}|c|c|c|c|c|}
\hline
Space group & n$^\circ$139 (D$_{4h}$)& n$^\circ$126 (D$_{4h}$) & n$^\circ$128 (D$_{4h}$) & n$^\circ$134 (D$_{4h}$) & n$^\circ$136 (D$_{4h}$) \\
\hline
Local Point symmetry at U & D$_{4h}$ & D$_{4}$ & C$_{4h}$ & D$_{2d}$ & D$_{2h}$ \\
\hline
\multirow{5}{3cm}{Possible A$_{2g}$ CEF transitions} & $\Gamma_1^+$ $\rightleftarrows$ $\Gamma_2^+$ & $\Gamma_1$ $\rightleftarrows$ $\Gamma_2$ & $\Gamma_1^+$ $\rightleftarrows$ $\Gamma_1^+$ & $\Gamma_1$ $\rightleftarrows$ $\Gamma_2$ & $\Gamma_1^+$ $\rightleftarrows$ $\Gamma_3^+$ \\
& $\Gamma_3^+$ $\rightleftarrows$ $\Gamma_4^+$ & $\Gamma_3$ $\rightleftarrows$ $\Gamma_4$ & $\Gamma_2^+$ $\rightleftarrows$ $\Gamma_2^+$ & $\Gamma_3$ $\rightleftarrows$ $\Gamma_4$ & \\
& $\Gamma_5^+$ $\rightleftarrows$ $\Gamma_5^+$ & $\Gamma_5$ $\rightleftarrows$ $\Gamma_5$ & $\Gamma_3^+$ $\rightleftarrows$ $\Gamma_4^+$ &  $\Gamma_5$ $\rightleftarrows$ $\Gamma_5$ & $\Gamma_2^+$ $\rightleftarrows$ $\Gamma_4^+$ \\
\cline{2-6}
& $\Gamma_6^+$ $\rightleftarrows$ $\Gamma_6^+$ & $\Gamma_6$ $\rightleftarrows$ $\Gamma_6$ & $\Gamma_5^+$ $\rightleftarrows$ $\Gamma_6^+$ & $\Gamma_6$ $\rightleftarrows$ $\Gamma_6$ & $\Gamma_5^+$ $\rightleftarrows$ $\Gamma_5^+$ \\
& $\Gamma_7^+$ $\rightleftarrows$ $\Gamma_7^+$ &  $\Gamma_7$ $\rightleftarrows$ $\Gamma_7$  & $\Gamma_7^+$ $\rightleftarrows$ $\Gamma_8^+$  & $\Gamma_7$ $\rightleftarrows$ $\Gamma_7$ &   \\
\hline
\end{tabular}
\caption{List of the CEF excitations localized on the Uranium atoms and Raman active in the A$_{2g}$ symmetry, for each space groups described in Table~\ref{tab1}. Some of these excitations are active in other symmetries (see the multiplication tables~\cite{koster_properties_1963}).}
\label{tab2}
\end{table*}

\bibliographystyle{apsrev4-1}
\bibliography{biblio}